\documentclass[preprint,12pt]{elsarticle}

\usepackage{graphicx}
\usepackage{amssymb}
 \usepackage{amsthm}
\usepackage{makecell}
\usepackage{tabularx}

\usepackage{lineno}

\journal{AAP, presented at the 2017 RSS Conference}

\begin{document}

\begin{frontmatter}


\title{Road safety of passing maneuvers: a bivariate extreme value theory approach under non-stationary conditions}



\author{Joana Cavadas}
\address{CITTA, Department of Civil Engineering, University of Coimbra\\ Rua Luís Reis Santos - P\'olo II, 3030-788 Coimbra, Portugal, joana.cavadas@uc.pt}

\author{Carlos Lima Azevedo}
\address{Department of Management Engineering, Technical University of Denmark\\ Anker Engelunds Vej 1, 2800 Kgs. Lyngby, climaz@dtu.dk}

\author{Haneen Farah}
\address{Department Transport and Planning, Delft University of Technology\\ Stevinweg 1, 2600 GA Delft, The Netherlands, h.farah@tudelft.nl}

\author{Ana Ferreira}
\address{Department of Mathematics, Instituto Superior Técnico. University of Lisbon\\ Av. Rovisco Pais 1, 1049-001 Lisbon, Portugal, anafh@tecnico.ulisboa.pt}

\begin{abstract}
Observed accidents have been the main resource for road safety analysis over the past decades. Although such reliance seems quite straightforward, the rare nature of these events has made safety difficult to assess, especially for new and innovative traffic treatments. Surrogate measures of safety have allowed to step away from traditional safety performance functions and analyze safety performance without relying on accident records. In recent years, the use of extreme value theory (EV) models in combination with surrogate safety measures to estimate accident probabilities has gained popularity within the safety community.
In this paper we extend existing efforts on EV for accident probability estimation for two dependent surrogate measures. Using detailed trajectory data from a driving simulator, we model the joint probability of head-on and rear-end collisions in passing maneuvers. In our estimation we account for driver specific characteristics and road infrastructure variables. We show that accounting for these factors improve the head-on collision probability estimation. This work highlights the importance of considering driver and road heterogeneity in evaluating related safety events, of relevance to interventions both for in-vehicle and infrastructure-based solutions. Such features are essential to keep up with the expectations from surrogate safety measures for the integrated analysis of accident phenomena.
\end{abstract}


\end{frontmatter}

%\linenumbers

\section{Introduction}
\label{S:1}

\subsection{Motivation}
\label{S:1.1}

Prediction of accidents has been a major topic in traffic safety for the last couple of decades. Despite the huge efforts that researchers have put in developing accident prediction models \cite{bonneson2010highway}, there is a great tendency in the last decades to develop new proactive methods for safety evaluation that are not based on accident records (\cite{archer2004methods}, \citep{kraay1985trautenfels}). Evaluating conflicts and risky situations between road users has been the main alternative and multiple methodologies can be found in the literature: the Swedish traffic conflict technique \citep{hyden1987development}, DOCTOR method \citep{kraay1985trautenfels}, and the use of surrogate safety measures \cite{archer2004methods}. The main challenge is the link between these measures and the number of accidents. \citeauthor{zheng2014freeway} (\citeyear{zheng2014freeway}) indicate that the validity of surrogate safety measures is usually determined by its correlation with accident frequency which is usually assessed using regression analysis. However, regression analysis still incorporates accident counts which are known to suffer from underreporting and quality issues, and thus this approach is limited. Besides, it is difficult to insure the stability of the accident-to-surrogate ratio, and this relationship also hardly reflects the physical nature of accident occurrence (\citep{zheng2014freeway}). Therefore, there is a need to develop an alternative approach to predict the number of accidents based on surrogate safety measures. \citeauthor{songchitruksa2006extreme} (\citeyear{songchitruksa2006extreme}) developed a new and more sophisticated approach based on the Extreme Value (EV) theory to estimate the frequency of accidents based on measured accident proximity.

\subsection{Extreme Value (EV) Approach}
\label{S:1.2}

The EV approach has three considerable advantages over the traffic conflict technique: (a) it abandons the assumption of fixed ratio converting the surrogate event frequency into accident frequency; (b) accident risk given the surrogate event is estimated based on the observed variability of accident proximity without using accident data; (c) the accident proximity measure precisely defines the surrogate event. 
The implicit assumption of the EV theory is that the stochastic behavior of the process being modeled is sufficiently smooth to enable extrapolation to unobserved levels \cite{de2007extreme}. In the context of road safety, the more observable traffic conflict events are used to predict the less frequent accidents, which are often unobservable in a short time period \citep{zheng2014freeway}. The field of EV theory, pioneered by \citeauthor{fisher1928limiting} (\citeyear{fisher1928limiting}), is a commonly applied theory in many fields, such as in meteorology, hydrology, finance \citep{zheng2014freeway} and very recently, road safety analysis \citep{songchitruksa2006extreme}. \citeauthor{songchitruksa2006extreme} (\citeyear{songchitruksa2006extreme}) used an EV approach to build up relationships between occurrences of right-angle accidents at urban intersections and frequency of traffic conflicts measured by using post-encroachment time. A major improvement of this study is that it links the probability of accident occurrence to the frequency of conflicts estimated from observed variability of accident proximity, using a probabilistic framework and without using accident records. The generic formulation of the application of EV to road safety analysis was then proposed by \citeauthor{tarko2012use} (\citeyear{tarko2012use}) and it was only very recently applied to other accident types and data sets (\citep{jonasson2014internal}, \citep{zheng2014freeway}). This formulation relies usually, but not exclusively, on time-based surrogate measures and estimates the probability of accident occurrence using the EV fitted distribution of such measures (see Appendix \ref{appendix A} for formulation details).

\subsection{Risk of Passing Maneuvers}
\label{S:1.3}

Passing maneuvers on two-lane roads (one lane per travel direction) carries several types of risks. The process of passing involves, synchronizing the vehicle’s speed with that of the vehicle in front, estimating the available gap on the opposite direction and evaluating its suitability to successfully perform the passing maneuver, and finally return to the main driving lane while keeping a sufficient safe gap from the passed vehicle, as well as, from the vehicle on the opposite direction. The gap from the passed vehicle at the end of the passing maneuver is termed in this study as ’THW’. It reflects the time headway between the front of the passed vehicle and the rear of the passing vehicle – a measure for rear-end and side-collisions with the passed vehicle. The gap from the vehicle on the opposite direction is termed in this study ‘TTC’ for time-to-collision between the passing and the opposite vehicle – a measure for head-on collisions. Both of these gaps are calculated at the end of the passing maneuver. In this study both measures will be used: the THW was calculated as the remaining distance between the passing and passed vehicle at the end of the passing maneuver divided by the driving speed of the passed vehicle, while the TTC was calculated as the remaining distance between the passing and opposing vehicle divided by the sum of their speeds. 

\subsection{Drivers’ Characteristics}
\label{S:1.4}

Several studies have shown that there are significant differences in passing behaviors among different drivers. \citeauthor{farah2011age} (\citeyear{farah2011age}) using a driving simulator found that gender and age have a significant impact on the passing behavior. She found that male drivers pass more frequently than female drivers. They also maintain smaller following time gaps from the front vehicle before initiating a passing maneuver and accept shorter gaps in the opposite traffic for passing. Younger drivers have significantly lower critical gaps and higher desired driving speeds compared to older drivers. They also keep smaller gaps from the front vehicle at the end of the passing maneuvers. These behaviors increase the risk of collisions. \citeauthor{llorca2013influence} (\citeyear{llorca2013influence}) reached similar conclusions using an instrumented vehicle. The authors found that young male drivers have shown a more aggressive behavior when passing compared to other groups of drivers. Passing times were around 1s lower than other drivers, while average speed difference was 4 km/h higher. \citeauthor{farah2007study} (\citeyear{farah2007study}) tested the significance of including driving styles in the passing behavior model, and found that drivers who are characterized by an anxious driving style and/or patient and careful driving style have larger critical gaps. \citeauthor{vlahogianni2012bayesian} (\citeyear{vlahogianni2012bayesian}) emphasize that the behaviors of young male and female drivers during passing maneuvers are different and this is because of differences in the process of scanning and evaluating available opportunities for passing. 
To summarize, the integration of drivers’ characteristics and driving styles in accident prediction is valuable and have the potential to contribute to understanding accident causation. Previous EV models did not account for such factors.

\section{Research Method}
\label{S:2}

The aim of this study is to test two different methods to estimate accident probability in passing maneuvers. The first approach analyzes the risk of individual types of accidents during passing maneuvers, including: (1) head-on collisions using the proximity measure of the minimum TTC to the vehicle in the opposite direction; (2) rear-end collisions using the proximity measure of the minimum THW measured from the front of the passed vehicle to the rear-end of the passing vehicle. The second approach aims to analyze the joint risk of colliding with the opposite or passed vehicle during passing maneuvers using the two surrogate safety measures (THW and TTC).

\subsection{Modeling Approach}
\label{S:2.1}

There are two families of EV distributions which follow two different approaches to sample extreme events: (1) the Generalized Extreme Value (GEV) distribution which is used in the block maxima or minima (BM) approach, in which maxima over blocks of time (or space) are considered; (2) the Generalized Pareto (GP) distribution which is used in the peaks over threshold (POT) approach \cite{coles2001introduction}, where all values above a certain threshold are used. In this paper both approaches are examined and compared. In this paper we focus our attention on the application of the BM approach for estimating the risk of a single type of accident (head-on or rear-end collision), while for estimating the risk of both types of collisions jointly, the bivariate distribution with copula approach was considered. It worth noting that the POT approach was also explored in this study. Yet, its results compared to the BM approach are discussed later in Section \ref{S:4} and, in more detail in Appendix \ref{appendix C}.

\subsection{The BM Approach}
\label{S:2.2}

In the GEV distribution the extreme events are sampled based on the BM approach. Following this approach, the observations are aggregated into fixed intervals over time or space, and then the extremes are extracted from each block by identifying the maxima in each single block. Mathematically, the standard GEV function is as follows \cite{zheng2014freeway}:

\begin{equation}
\label{eq:gev}
G(x)=exp \left( - \left[ 1+ \xi \left( \frac{x-u}{\sigma} \right) \right] ^{-1/\xi} \right)
\end{equation}

If $X_1, X_2, \dots, X_n$ is a set of independently and identically distributed random observations with unknown distribution function $F \left( x \right) = Pr {\left({X}_{i} \leq x \right)}$, the linearly normalized maximum ${M}_{n} =max \left[ {X}_{1}, {X}_{2}, \dots, {X}_{n} \right]$ will converge to a GEV distribution when $n\to\infty$. Three parameters identify this distribution: the location parameter, $-\infty < u(z) < \infty$; the scale parameter, $\sigma>0$; and the shape parameter, $-\infty < \xi < \infty$. If the shape parameter, $\xi$ , is positive, then his would yield the Frechet Cumulative Distribution Function (CDF) with a finite lower endpoint, $(u-\sigma/\xi)$, if $\xi$  is negative, this will yield the (reversed) Weibull CDF with finite upper endpoint $({u-\sigma}/ {\left| \xi \right|})$, and if $\xi=0$ this yields the Gumbel CDF. In a non-stationary BM model several factors, $z$, can be included in the location parameter to account for their impact on the probability of the extreme events, i.e. $u(z)$. More details on the GEV properties can be found in \cite{tarko2012use}.

In this study, all EV models’ were estimated using the maximum likelihood method (ML) in \texttt{R} (v3.0.3) using the \texttt{exTremes} and \texttt{evd} packages \citeauthor{gilleland2011new} (\citeyear{gilleland2011new}). Details on the statistical properties of EV can be found in (\cite{coles2001introduction}, \cite{de2007extreme}), and on the theoretical background of its applicability for surrogate (road) safety analysis in \cite{tarko2012use}.

\subsection{Bivariate EV distribution}
\label{S:2.3}

In some applications, the study of accident probability using multivariate distributions is of interest. Traditionally, single surrogate safety measures are used to estimate a single type of events. However, it is expected that in some of the complex accident phenomena, multiple pre-accident events can play an important role in a potential accident. Passing maneuvers are a typical case where both the opposite and passed vehicles are key stimulus during driver’s decision making.
Given a bivariate random sample $(X_1,Y_1),\dots,(X_n,Y_n)$, much of extreme value theory is concerned with the limiting behavior of a suitable normalization of the component wise maxima $(M_{1,n},M_{2,n}$, where ${M}_{1,n} = max (X_1,\dots,X_n)$ and ${M}_{1,n} = max (Y_1,\dots,Y_n)$. More precisely, it is assumed that there exists a non-degenerate bivariate distribution function $L$ such that, as $n\to\infty$:

\begin{equation}
\label{eq:bivar}
Pr \left\{ {\frac{M_{1,n}-b_{1,n}}{a_{1,n}} \leq x, \frac{M_{2,n}-b_{2,n}}{a_{2,n}} \leq y} \right\} \to L(x,y)
\end{equation}

for sequences $a_{j,n}>0, b_{j,n} \in R,j=1,2$  \cite{caperaa2000estimation}. To analyze separately the behavior of the marginals and the dependence structure of the distribution, it is convenient to write $L(x,y)=C\{F(x),G(y)\}$ in terms of univariate extreme value margins $F$ and $G$, and a "copula" $C$ \cite{caperaa2000estimation} defined for all $0 \leq w,v \leq 1$ by:

\begin{equation}
\label{eq:copula}
C \left ( w,v \right ) = Pr \lbrace F \left ( X \right ) \leq w, G \left ( Y \right ) \leq v \rbrace = exp \left [ log { \left (w v \right ) } {A}^{*} \left \{ \frac{log { ( w ) } } { log ⁡(w v)} \right\} \right ]
\end{equation}

where ${A}^{*}$ is a convex function on $\left [ 0,1 \right ]$ such that $max (t,1-t) \leq {A}^{*}(t) \leq 1$ for all $0 \leq t \leq 1$ \cite{pickands1981multivariate}. Given this representation and except for the margins, the bivariate extreme value distribution $L$ for component wise maxima is characterized by a one-dimensional function ${A}^{*}$. Common used functions are the logistic, asymmetric logistic, Husler-Reiss, negative logistic, asymmetric negative logistic, bilogistic, negative bilogistic, Coles Tawn, and asymmetric mixed models. Further information on the statistical properties of the estimation methods that have been developed in the context of bivariate EV can be found in (\cite{caperaa2000estimation}, \cite{pickands1981multivariate}). In this study, bivariate models’ were estimated using the ML using the \texttt{VineCopula} and \texttt{evd} packages in \texttt{R}.

\subsection{Bivariate distributions with Copula method to model dependence}
\label{S:2.4}

We focus our attention on the bivariate distribution with a copula method due to the expectable existence of dependence (not necessary linear) between the two surrogate safety measures. A copula is a multivariate distribution whose margins are all uniform over (0,1). For a 2-dimensional random vector $U$ on the unit cube, a copula can be defined as:

\begin{equation}
\label{eq:copula2}
C \left ( u_1,u_2 \right ) = Pr \lbrace U_1 \leq u_1, U_2 \leq u_2 \rbrace
\end{equation}

The copula not only provides a structure for the dependence between the two variables but also reveals itself to be invariant under strictly monotone transformations. The Sklar’s Theorem (1959) ensures that it is possible to estimate a multivariate distribution by separately estimating the marginal distributions and the copula function $C$. In this sense, let $F$ be the 2-dimensional distribution function of the random vector $X$ with margins $F_1,F_2$, then the copula $C$ is such that for all vector $x$ the equality $F(x)=C(F_1(x_1),F_2(x_2))$ holds, where $C(\cdot)$ is unique if the marginal distributions are continuous. 
The two most frequently used copula families are elliptical copulas and Archimedean copulas. More details on these copula families can be found in \cite{fang2017symmetric}, \cite{nelsen2007introduction} and \cite{genest1993statistical}. To assess if a given copula is well fitted to the data under analysis, a goodness-of-fit test is performed based on statistics such as the rank-based versions of the Cramer-von Mises or the Kolmogorov-Smirnov. An example of goodness-of-fit testing overview are given in \citeauthor{berg2009copula} (\citeyear{berg2009copula}).

\subsection{Data Collection}
\label{S:2.5}
The data for this study was obtained from a driving simulator experiment developed by \citeauthor{farah2009risk} (\citeyear{farah2009risk}) for modelling drivers’ passing behavior on two-lane rural highways. In this experiment the STISIM \cite{rosenthal1999stisim} driving simulator was used. STISIM is a fixed-base interactive driving simulator, which has a 60° horizontal and 40° vertical display. The driving scene was projected onto a screen in front of the driver with a rate of 30 frames per second. A total of 16 simulator scenarios were designed in order to have a better understating of how different infrastructure and traffic related factors affect drivers’ passing behavior. The 16 different scenarios are the result of an experimental design that included 4 factors in 2 levels, which are: the speed of the front vehicle (60 or 80 km/h), the speed of the opposite vehicle (65 or 85 km/h), the opposite lane traffic volume (200 or 400 veh/h), and the road curve radius (300-400 or 1500-2500 m). However, all the scenarios were composed of 7.5 km of two-lane rural highway section with no intersections, and good weather conditions. Each driver drove 4 scenarios out of the 16 scenarios which were selected following a partial confounding method that was adopted \citeauthor{hicks1999fundamental}. A more detailed information about this experiment can be found in \cite{farah2013modeling}, \cite{farah2009risk}.
A total of 100 drivers (64 males and 36 females) with at least 5 years of driving experience participated in the driving simulator experiment on a voluntary base. 67 drivers are with an age between 22 and 34 years old, 20 drivers with an age between 35 and 49 years old, and the remaining 12 with an age between 50 and 70 years old.
Prior to participating in the driving simulator experiment each driver filled a questionnaire composed of two parts: the first part included questions on the driver personal characteristics (including questions such as: gender, age, and driving experience), while the second part included the Multidimensional Driving Style Inventory (MDSI) developed by \citeauthor{taubman2004multidimensional} (\citeyear{taubman2004multidimensional}). The MDSI is a 6-point scale, which consists of 44 items that are used to characterize four factors that represent different driving styles: (1) Reckless and careless driving style, which refers to deliberate violations of safe driving norms, and the seeking of sensations and thrills while driving. It characterizes persons who drive at high speeds, race in cars, pass other cars in no-passing zones, and drive while intoxicated, probably endangering themselves and others; (2) Anxious driving style, which reflects feelings of alertness and tension as well as ineffective engagement in relaxing activities during driving; (3) Angry and hostile driving style, which refers to expressions of irritation, rage, and hostile attitudes and acts while driving, and reflects a tendency to act aggressively on the road, curse, blow horn, or “flash” to other drivers, and (4) Patient and careful driving style, which refers to planning ahead, attention, patience, politeness, and calmness while driving, as well as obedience to traffic rules. Factor scores were calculated for each respondent on each of these four driving styles.

\section{Results and Analysis}
\label{S:3}

The data set from the driving simulator experiment resulted in a total of 1287 completed passing maneuvers, 9 head-on collisions and 2 rear-end collisions. The detailed vehicle movement from the simulator was processed to obtain the two surrogate safety measures of interest at the end of passing maneuvers: the TTC with the opposing vehicles and the THW between the passed and passing vehicles.

\subsection{Univariate Model}
\label{S:3.1}

In the univariate model, a separate distribution was fitted to the minimum TTC and THW measurements resulting from the1287 passing maneuvers. In the GEV approach, each passing maneuver is represented by one block for which we take its minima for each of the surrogate safety measures considered. Note that accident observations were not used in the estimation procedure. GEV (and GP) models pertain to continuous random variables that give zero mass to any real value, hence to zero. But accidents do happen and can be recorded with zero value with positive mass. The continuous random variables do not take such values into account and, thus, recorded accidents were not considered in the estimation dataset. 

\begin{figure}[h!]
\centering\includegraphics[width=1\linewidth]{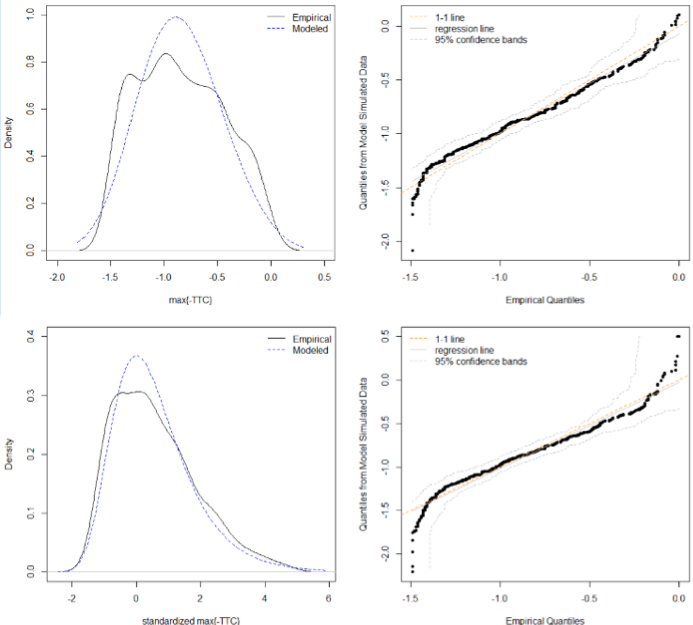}
\caption{Probability Density plots (left) and simulated QQ-plots (right) for the stationary BM model (top) and the best non-stationary BM model (Model \#1, bottom)}
\label{img:figure_1}
\end{figure}

Aiming at estimating the probability of a head-on collision for a single passing maneuver, the minimum TTC was considered as a head-on accident surrogate measure. The data was then filtered to account only for values smaller than 1.5s (\cite{hyden1987development}, \cite{jonasson2014internal}, \cite{vogel2003comparison}), leading to a total of 463 observations. Knowing that 9 maneuvers ended with actual head-on collisions, the empirical probability of a head-on collision in a passing maneuver given that a critical TTC (i.e. TTC lower than 1.5 s), is 9/(463+9)=0.0191, with 95\% binomial confidence interval (0.0067,0.0314). Note that different filtering conditions were also tested in estimation (see Appendix \ref{appendix C}). 
We start with an existing stationary BM model developed by \citeauthor{farah2017safety} (\citeyear{farah2017safety}). The authors estimated that the parameters of the univariate GEV cumulative distribution function are $\hat{\mu} =-0.993(0.0212)$ , $\hat{\sigma} = 0.0383(0.0163)$ and $\hat{\xi}=-0.236(0.0500)$. Figure \ref{img:figure_1}. presents the probability density function of the empirical and modeled negated TTC (\texttt{upper left}) and the simulated QQ plot (\texttt{upper right}). This model was then upgraded by the authors to a non-stationary BM model. They concluded that the covariates ‘\textit{passinggap}’, ‘\textit{tailgatetp}’, ‘\textit{speedfront}’, ‘\textit{curvature}’ as defined below, related to the infrastructure and traffic, significantly contribute to the prediction of the probability of a head-on-collision during a passing maneuver. While the covariate ‘\textit{speedpv}’ was not found to be significant. Variables related to drivers’ personal characteristic (\textit{gender}, \textit{age}, and \textit{driving style}) were not tested. In this study, we will test whether drivers’ personal characteristic significantly contribute to the model in addition to the traffic and road variables. The variables are defined as following:

\begin{enumerate}
    \item \textit{passinggap}: time gap between two opposite vehicles at the time the subject meets the opposite vehicle;
    \item \textit{tailgatetp}: time gap between the subject vehicle and the front vehicle at the moment of start passing (s);
    \item \textit{speedfront}: speed of the front vehicle at the moment of start passing (m/s);
    \item \textit{curvature}: road curvature (1/m);
    \item \textit{speedpv}: speed of the passing vehicle (m/s);
    \item \textit{gender}: gender of the driver (1-male; 0-female);
    \item \textit{age}: categorical variable, with ranges 22-34; 35-49 and 50-70;
    \item \textit{drivingstyle}: angry \& hostile; anxious; reckless \& careless; patient \& careful (Taubman-Ben-Ari et al., 2004);
\end{enumerate}

A set of a non-stationary models considering different combinations of covariates were estimated. Table \ref{tab:table_1} presents the four best non-stationary models with a range of likelihood ratio p-value between $2.773 \times 10^{-9}$ and $1.931 \times 10^{-8}$ (Model \#1 to \#4). The estimated likelihood ratio tests are shown in Table \ref{tab:table_2}. The previously estimated non-stationary model by \citeauthor{farah2017safety} (\citeyear{farah2017safety}), presented as Model \#0, is used as a benchmark for assessing the performance of the other models. 
Analyzing the results presented in Table \ref{tab:table_1}, it is concluded that the inclusion of \textit{gender} (model \#1) improves the accuracy of the model when compared to the non-stationary model (\#0). The significance of this variable is given by the p-value of the likelihood ratio test, which is equal to $0.023$ as presented in Table \ref{tab:table_2}, with 95\% confidence level. 

\begin{table}[h]
\small
\centering
\begin{tabular}{l l l l l l}
\Xhline{2\arrayrulewidth}
\textbf{\makecell{Non-stationary \\ model}} & \textbf{\#0} & \textbf{\#1} & \textbf{\#2} & \textbf{\#3} & \textbf{\#4}\\
\Xhline{2\arrayrulewidth}
$\hat{\mu}_0$ &
\makecell[l]{-1.045 \\ \footnotesize{(0.137)}} &
\makecell[l]{-0.983 \\ \footnotesize{(0.139)}} &
\makecell[l]{-0.927 \\ \footnotesize{(0.145)}} &
\makecell[l]{-0.953 \\ \footnotesize{(0.145)}} &
\makecell[l]{-1.107 \\ \footnotesize{(0.139)}} \\
\Xhline{0.5\arrayrulewidth}

$\hat{\mu}_1$(\textit{speedFront}) &
\makecell[l]{0.024  \\ \footnotesize{(0.006)}} &
\makecell[l]{0.026  \\ \footnotesize{(0.006)}} &
\makecell[l]{0.027  \\ \footnotesize{(0.006)}} &
\makecell[l]{0.025  \\ \footnotesize{(0.006)}} &
\makecell[l]{0.027  \\ \footnotesize{(0.006)}} \\
\Xhline{0.5\arrayrulewidth}

$\hat{\mu}_2$(\textit{tailgatetp}) &
\makecell[l]{0.002  \\ \footnotesize{(0.002)}} &
\makecell[l]{0.003  \\ \footnotesize{(0.002)}} &
\makecell[l]{0.003  \\ \footnotesize{(0.002)}} &
\makecell[l]{0.003  \\ \footnotesize{(0.002)}} &
\makecell[l]{0.003  \\ \footnotesize{(0.002)}} \\
\Xhline{0.5\arrayrulewidth}

$\hat{\mu}_3$(\textit{passinggap}) &
\makecell[l]{-0.022  \\ \footnotesize{(0.004)}} &
\makecell[l]{-0.023  \\ \footnotesize{(0.004)}} &
\makecell[l]{-0.023  \\ \footnotesize{(0.004)}} &
\makecell[l]{-0.023  \\ \footnotesize{(0.004)}} &
\makecell[l]{-0.023  \\ \footnotesize{(0.004)}} \\
\Xhline{0.5\arrayrulewidth}

$\hat{\mu}_4$(\textit{curvature}) &
\makecell[l]{-33.653  \\ \footnotesize{(13.519)}} &
\makecell[l]{-34.304  \\ \footnotesize{(13.419)}} &
\makecell[l]{-34.068  \\ \footnotesize{(13.403)}} &
\makecell[l]{-34.090  \\ \footnotesize{(13.488)}} &
\makecell[l]{-34.139  \\ \footnotesize{(13.397)}} \\
\Xhline{0.5\arrayrulewidth}

$\hat{\mu}_5$ (\textit{Gender}) &
- &
\makecell[l]{-0.097  \\ \footnotesize{(0.042)}} &
\makecell[l]{-0.080  \\ \footnotesize{(0.043)}} &
- &
-\\
\Xhline{0.5\arrayrulewidth}

$\hat{\mu}_6$(\textit{Angry}\&\textit{Hostile}) &
- &
- &
\makecell[l]{-0.021  \\ \footnotesize{(0.016)}} &
\makecell[l]{-0.029  \\ \footnotesize{(0.015)}} &
-\\
\Xhline{0.5\arrayrulewidth}

$\hat{\mu}_7$(\textit{F2234}) &
- &
- &
- &
- &
\makecell[l]{0.116  \\ \footnotesize{(0.044)}}\\
\Xhline{0.5\arrayrulewidth}

$\hat{\sigma}$ &
\makecell[l]{0.364  \\ \footnotesize{(0.015)}} &
\makecell[l]{0.362  \\ \footnotesize{(0.014)}} &
\makecell[l]{0.361  \\ \footnotesize{(0.014)}} &
\makecell[l]{0.360  \\ \footnotesize{(0.014)}} &
\makecell[l]{0.361  \\ \footnotesize{(0.014)}} \\
\Xhline{0.5\arrayrulewidth}

$\hat{\epsilon}$ &
\makecell[l]{-0.219  \\ \footnotesize{(0.042)}} &
\makecell[l]{-0.217  \\ \footnotesize{(0.041)}} &
\makecell[l]{-0.216  \\ \footnotesize{(0.041)}} &
\makecell[l]{-0.216 \\ \footnotesize{(0.041)}} &
\makecell[l]{-0.217  \\ \footnotesize{(0.041)}} \\
\Xhline{2\arrayrulewidth}

Neg. LL &
208.65 &
206.06 &
205.22 &
206.89 &
205.24\\
\Xhline{2\arrayrulewidth}

\end{tabular}
\caption{Estimation results (and std. error) of the non-stationary BM approach for head-on collisions
Non-stationary model}
  \label{tab:table_1}
\end{table}

\begin{table}[h]
\small
\centering
\begin{tabular}{l l l l l l}
\Xhline{2\arrayrulewidth}
\textbf{Model} & \textbf{\#0} & \textbf{\#1} & \textbf{\#2} & \textbf{\#3} & \textbf{\#4}\\
\Xhline{2\arrayrulewidth}
\#0 & - \\				
\#1	& 5.189 \footnotesize{(0.023)}	& & - \\		
\#2	& 6.872 \footnotesize{(0.032)}	& 1.683 \footnotesize{(0.194)} & -	\\	
\#3	& 3.527 \footnotesize{(0.060)}	& -1.662 \footnotesize{(1.000)} & 3.345 (0.067) & -\\	
\#4	& 6.832 \footnotesize{(0.008)}	& 1.644  \footnotesize{($2.2\times10^{-16}$)}	& 0.039 \footnotesize{(0.843)}	& 3.306 \footnotesize{($2.2\times10^{-16}$)}	& - \\
\Xhline{2\arrayrulewidth}

\end{tabular}
\caption{Likelihood Ratio Test (and p-value) for the non-stationary BM models for head-on collisions}
  \label{tab:table_2}
\end{table}

The contribution of the variables representing driving styles (\textit{Angry}\&\textit{Hostile}, \textit{Anxious}, \textit{Reckless}\&\textit{Careless} and \textit{Patient}\&\textit{Careful}) was tested considering all the possible combinations of these variables besides the ones included in model \#1. Comparisons between the different models were based on the likelihood ratio test. This procedure resulted in the inclusion of one driving style, \textit{Angry}\&\textit{Hostile}, as presented in model \#2. Analyzing the correlation between the different driving styles and the sociodemographic variables, a small but significant sample correlation of 0.29 was found between \textit{Angry}\&\textit{Hostile} and \textit{Gender}. For modelling purposes, and in order to test which variable among the two has a larger influence, the variable \textit{Gender} was excluded from model \#2, creating model \#3. Comparing the results of models \#1 to \#3, it is concluded that the model that only includes \textit{Gender} (model \#1) has a better fit based on the p-value of the likelihood ratio test. Although \textit{Angry}\&\textit{Hostile} could have higher explanatory power in other samples, a reason to prefer this model is the simplicity of collecting data on driver gender compared to drivers’ driving styles, which requires the completion of the MDSI survey.
Aligned with the conclusions achieved by \citeauthor{farah2011age} (\citeyear{farah2011age}) and \citeauthor{llorca2013influence} (\citeyear{llorca2013influence}) regarding the impact of age, where it was found to improve the accuracy of the model when compared to the stationary model but turned out to have a non-significant contribution if gender is also included. After several attempts, we included the interaction variable between \textit{Gender} (female drivers) with age (range 22-34), \textit{F2234}, and the final model (model \#4) is shown in Table \ref{tab:table_1}. This model considers a new variable that takes 1 if the driver is a female with age range between 22 and 34, and zero otherwise. 
To estimate the probability of a head-on collision along with the conclusion about which model is the one with the better fit (models \#1 and \#4), two different approaches were considered. The first approach considers that the location parameter value is calculated using the covariates from the data, achieving the estimated probabilities of 0.0195 and 0.0198 for models 1 and 4, respectively, with 95\% confidence level (0.0192; 0.0198) and (0.0195; 0.0201), respectively. These confidence intervals of estimation were computed assuming a normal distribution under regular parameters’ conditions, a simulation experiment size of $1\times10^{6}$ and its simulated distribution quantiles. The second approach considers the estimation of the location parameters based on the estimation dataset, where normal distributions with means (standard deviations) of -0.989 (0.123) and -0.988 (0.125), for models \#1 and \#4, respectively were considered. The Kolmogorov-Smirnov test statistic of 0.0444 and 0.0479, respectively was achieved. This procedure simulates the values 0.0197 and 0.0202 for the probabilities of head-on collisions of models \#1 and \#4, respectively, with 95\% confidence interval of (0.01939, 0.0199) and (0.0199, 0.0205).  Comparing the probabilities of these two methods with the probability for a head-on collision assuming a near head-on collision in a passing maneuver of 0.0191, results in model \#1 give a slightly better estimation compared to model \#4 and the estimation performance is not significantly deteriorated.
According to the results of model \#1 presented in Table \ref{tab:table_1}, if the speed of the front vehicle (\textit{speedfront}) increases, or if drivers start their passing maneuver from a larger gap from the front vehicle (\textit{tailgatetp}), the negated TTC increases (corresponding to a decrease in the TTC). These are logical results since it is more difficult to end the passing maneuver if the front vehicle has a higher driving speed. Similarly, starting the passing maneuver from a larger gap from the front vehicle results in a longer time to finish the maneuver and consequently smaller TTC. If the passing gap (\textit{passinggap}) that is accepted is larger, or the curvature of the road (\textit{curvature}) is larger, the negated TTC is lower and the TTC is higher. This shows that drivers adapt their behavior in a passing maneuver if the road is too complex (i.e. sharp curves). Finally, male drivers have smaller TTC. This result is supported by previous studies (\cite{farah2007study}, \cite{vlahogianni2012bayesian}), where it was found that male drivers usually drive faster, have shorter passing gaps, and conduct a higher number of passing manoeuvers when compared to females. 

The probability density function of the empirical and modeled standardized \footnote{For non-stationary models, it is common practice to transform the data to a density function that does not depend on the covariates, using the following function $Z_1=-log\left(1+\frac{\epsilon}{\sigma}\left(X_i-\mu_i\right)\right)^{-\frac{1}{\epsilon}}$ \cite{gilleland2011new}.} maximum negated TTC and the simulated QQ-plot for the best non-stationary BM model (model \#1) are shown in Figure \ref{img:figure_1}. From these figures it can be concluded that the modeled GEV distribution has satisfactory fitting results to the empirical data.

\subsection{Rear-end collisions}
\label{S:3.2}

In order to estimate the probability of rear-end collisions for a single passing maneuver, the headway between the front passed vehicle and the passing vehicle at the end of the passing maneuver (THW) is used as accident surrogate. Similar to what was developed to calculate the probability of a head-on collision for a single passing maneuver, the minimum THW should be smaller than a limit to be useful as an accident surrogate. Based on the literature, this value varies between 0.6s \cite{vogel2003comparison} and 2.0s (\cite{evans1982accident}, \cite{vogel2003comparison}). Considering these thresholds, several BM stationary models were developed. Based on the estimates presented in \ref{appendix C}, the value of 2.0s was selected.

With a total of 492 observations with a THW smaller than 2.0s and knowing that 2 rear-end collisions occurred, the theoretical probability of a rear-end collision was calculated as $2/(492+2)= 0.00405$, with a 95\% binomial confidence interval (-0.00155, 0.00964).

The estimation of the stationary BM for the model of the negated values of the THW as carried out using the Gumbel distribution as the stable region for the shape parameter around the 2.0s THW filter resulted in $\hat{\xi}\approx0$. Further, as explored in \ref{appendix C}, we use the normalized dataset 
$\tilde{X_{i}}=-X_{i}-max \left( -\{X_{j}\}_{j=1}^n \right)$, $X_i$ being the THW for observation $i$, that was proven to provide a better prediction performance. Thus, we obtained the parameters  $\hat{\mu}=-1.456(0.0121)$ , $\hat{\sigma}=0.256(0.0093)$ and $\hat{\xi}=0$. The density function of the empirical and modeled negated THW and the simulated QQ plot are shown in Figure \ref{img:figure_2}. As explored in \ref{appendix C}, it is worth noting the exploration of the POT estimation for modelling rear-end collisions. Due to the small sample and high variance of the surrogate measure at stake (low THW) the non-normalized BM estimation resulted in positive shape parameter for low filtering conditions. Yet, the POT approach was able to perform well in estimation and prediction with thresholds between 1.2 and 1.5s. As this method seems to provide a better fit to this particular phenomenon at stake, especially under small samples, it may be of interest to explore in future research the combination of different marginal distribution in the estimation of bi-variate models for overpassing maneuvers.
Using the fitted Gumbel distribution to the normalized data, the estimated probability of this stationary model is 0.00334 with 95\% confidence interval (0.00317, 0.00340). This interval was computed assuming a normal distribution under regularity conditions of the parameters, simulating an experiment with a size of $1\times10^{6}$ and its simulated distribution quantiles. This estimated probability is relatively close to the empirical probability of 0.00405.

Notwithstanding, the passing maneuver may be affected by specific passing conditions, such as speeds of the vehicles surrounding the subject vehicle. Therefore, several linear combinations of covariates were tested according to a non-stationary BM model approach. This process was conducted in a similar way to the model developed to estimate the probability of a head-on-collision.

Taking this into account, we start with the non-stationary BM model \#0, which includes only the best covariates related with the maneuver and the environment: passing vehicle speed (\textit{speedpv}) and the passing gap time (\textit{passinggap}). Testing this non-stationary model against the stationary one through the likelihood ratio test, a p-value of 0.0002 is achieved with a direct value of 17.508 (note, 2 degrees of freedom). We then estimate the probability of a rear-end collision for a single passing maneuver (see \ref{appendix B}). This distribution with a mean of -1.454, a standard deviation of 0.0232 and a Kolmogorov-Smirnorv test statistic of 0.025, lead to a simulated rear-end collision probability of 0.00339 with 95\% confidence interval (0.00328, 0.00351), resulting in a slightly better estimation than the stationary model. The probability density plot as well as the QQ-plot for model \#0 are shown in Figure \ref{img:figure_2}.

\begin{table}[h]
\small
\centering
\begin{tabular}{l l l l}
\Xhline{2\arrayrulewidth}
\textbf{\makecell{Non-stationary \\ model}} & \textbf{\#0} & \textbf{\#1} & \textbf{\#2} \\
\Xhline{2\arrayrulewidth}
$\hat{\mu}_0$ &
\makecell[l]{-1.42 \\ \footnotesize{(0.0626)}} &
\makecell[l]{-1.37 \\ \footnotesize{(0.0653)}} &
\makecell[l]{-1.42 \\ \footnotesize{(0.0694)}} \\
\Xhline{0.5\arrayrulewidth}

$\hat{\mu}_1$(speedFront) &
\makecell[l]{0.00587 \\ \footnotesize{(0.00193)}} &
\makecell[l]{0.00686  \\ \footnotesize{(0.00198)}} &
\makecell[l]{0.00673  \\ \footnotesize{(0.00197)}} \\
\Xhline{0.5\arrayrulewidth}

$\hat{\mu}_3$(passinggap) &
\makecell[l]{-0.00916  \\ \footnotesize{(0.00263)}} &
\makecell[l]{-0.0101  \\ \footnotesize{(0.00267)}} &
\makecell[l]{-0.0107  \\ \footnotesize{(0.00271)}} \\
\Xhline{0.5\arrayrulewidth}

$\hat{\mu}_5$ (Gender) &
- &
\makecell[l]{-0.0590  \\ \footnotesize{(0.0282)}} &
\makecell[l]{-0.0753   \\ \footnotesize{(0.0296)}} \\
\Xhline{0.5\arrayrulewidth}

$\hat{\mu}_6$(Angry\&Hostile) &
- &
- &
\makecell[l]{0.0185  \\ \footnotesize{(0.00992)}}\\
\Xhline{0.5\arrayrulewidth}

$\hat{\sigma}$ &
\makecell[l]{0.254  \\ \footnotesize{(0.00914)}} &
\makecell[l]{0.253  \\ \footnotesize{(0.00910)}} &
\makecell[l]{0.252  \\ \footnotesize{(0.00906)}} \\
\Xhline{2\arrayrulewidth}

$\hat{\epsilon}$ &
0 &
0 &
0 \\
\Xhline{2\arrayrulewidth}

Neg. LL &
104.09 &
101.98 &
100.26 \\
\Xhline{2\arrayrulewidth}

\end{tabular}
\caption{Estimation results of the non-stationary BM approach for rear-end collisions}
  \label{tab:table_3}
\end{table}

\begin{table}[h]
\small
\centering
\begin{tabular}{l l l l}
\Xhline{2\arrayrulewidth}
\textbf{Model} & \textbf{\#0} & \textbf{\#1} & \textbf{\#2} \\
\Xhline{2\arrayrulewidth}
\#0 & - \\				
\#1	& 4.23 \footnotesize{(0.0398)}	& & - \\		
\#2	& 7.657 \footnotesize{(0.0218)}	& 3.43 \footnotesize{(0.0640)} & -	\\	
\Xhline{2\arrayrulewidth}
\end{tabular}
\caption{Likelihood Ratio Test (and p-value) for the non-stationary BM models for rear-end collisions}
  \label{tab:table_4}
\end{table}

From all models including driver’s characteristics, Table \ref{tab:table_3} presents the two best models for rear-end collisions, which actually relied in the same key co-variates as the head-on collision estimation results. While age related co-variates did not bring any improvement in the estimation results, Gender and \textit{Angry}\&\textit{Hostile} co-variates managed to improve it. Again, the significance of these variable is given by the p-value of the likelihood ratio tests presented in Table \ref{tab:table_4}. We note again the correlation found between these two variables and as previously discussed, followed with model \#1 in the rest of this paper.

Similarly to the head-on estimates, two different approaches were considered for probability estimation. The first approach considers that the location parameter value is calculated using the covariates from the data. The second approach considers the estimation of the location parameter distribution based on the estimation dataset. The estimated probabilities of 0.00333 and 0.00337, respectively, with 95\% confidence level (0.00322; 0.00345) and (0.00326; 0.00349) were obtained. All confidence intervals of estimation were computed assuming a normal distribution under regular parameters’ conditions, a simulation experiment size of $1\times10^{6}$ and its simulated distribution quantiles. 

\begin{figure}[h!]
%\begin{subfigure}
\centering\includegraphics[width=1\linewidth]{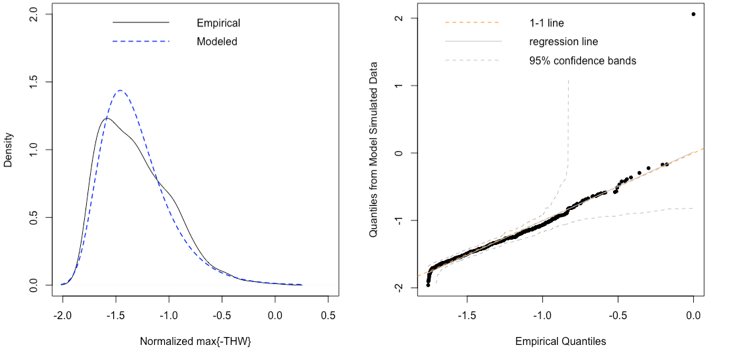}
%\end{subfigure}
%\begin{subfigure}
\centering\includegraphics[width=1\linewidth]{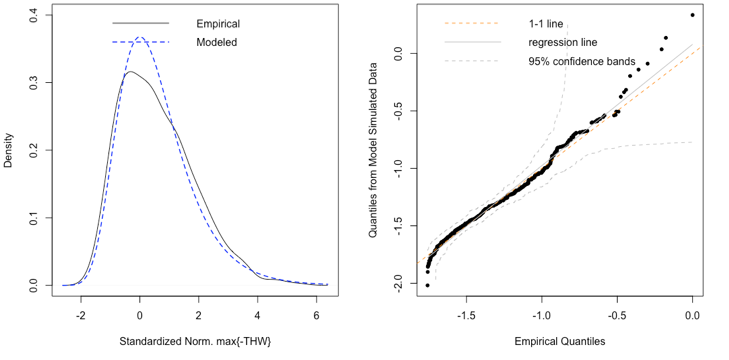}
%\end{subfigure}

\caption{Probability Density plots (left) and simulated QQ-plots (right) for the stationary BM model (top) and the best non-stationary BM model (Model \#1, bottom)}
\label{img:figure_2}
\end{figure}

\section{Bivariate Model}
\label{S:3.2}
It is aimed to estimate a measure of risk that not only takes into account the possibility to collide with the opposite vehicle but also with the passed vehicle. Taking into account that performing a passing maneuver requires a split of attention by the driver regarding its location relative to the surrounding vehicle, it is assumed that the dependence between the TTC and the THW is unknown. Furthermore, an integrated analysis is possible to be developed if, and only if, a relationship of dependence can be found between TTC and THW. When examining the correlation between these two variables with using the whole dataset, a Pearson-correlation value of 0.186 was found. This value shows the lack of linear correlation between the time-to-collision measure and the time head-way measure. However, this does not mean that TTC and THW are independent \cite{embrechts2002correlation}. To further examine potential correlation, the Kendall’s rank correlation tau was computed and found to be significantly greater than zero, indicating the existence of dependence between TTC and THW ($\tau = 0.192$, p-value $ <2.2\times10^{-16}$). This statement is corroborated by the independence test Global Cramer-von Mises, where a significant p-value close to zero (p-value=0.000499) gives strong evidence against the null hypothesis of independence.

We thus start by estimating a stationary bivariate model. For this exploratory analysis, we estimate the bivariate distribution considering the stationary univariate BM distributions for each variable as the upper tail margins distributions. This integration estimates the probability of an accident conditioned on TTC and THW being their thresholds (1.5s and 2.0s, respectively). By following this procedure, we assume that to perform the estimation for the remaining TTC and THW values, other distributions for the margins should be analyzed and fitted (e.g., gamma distributions or Gumbel distribution). This is mainly due to the reported non-extreme value distribution of safety measures beyond surrogate safety analysis conditions.

Using the R package \texttt{evd}, we explore the bivariate logistic distribution function \cite{gumbel1960distributions} with parameter $r$ for Eq. 5 above. This is a special case of the bivariate asymmetric logistic model where complete dependence is obtained in the limit as $r$ approaches zero (note that independence is obtained when $r = 1$). The results are obtained are presented in Table \ref{tab:table_5}. A tail dependence of 0.1783 and an estimated probability of accident of 0.0141 was obtained. The empirical probability for comparison was calculated by knowing that a maximum of 5 (4 head-on and 1 rear-end collisions) out of the total 11 collisions had the other surrogate safety measure value below its threshold (i.e., if head-on collision, then THW was below 2.0 s, and if rear-end collision, then the TTC was below 1.5 s), and that the sample of size is given by the number of collisions plus the 256 observations where both TTC and THW were below 1.5 and 2.0, respectively. Therefore, the empirical collision probability is 0.0191 (0.0025, 0.0358).

\begin{table}[h]
\small
\centering
\begin{tabular}{l l l l}
\Xhline{2\arrayrulewidth}
 &  \textbf{TTC} &  \textbf{THW}\\
$\hat{\mu}$ & -0.886 \footnotesize{(0.0309)} & -1.417 \footnotesize{(0.0207)}\\
$\hat{\sigma}$ & 0.431 \footnotesize{(0.0244)} & 0.280 \footnotesize{(0.0156)}\\
$\hat{\xi}$ &  -0.417 \footnotesize{(0.0600)} & -0.00083 \footnotesize{(0.0596)}\\
\Xhline{2\arrayrulewidth}
r	& 0.865 \footnotesize{(0.0390)}\\
AIC	& 419.197\\
\Xhline{2\arrayrulewidth}
\end{tabular}
\caption{Estimation results for the stationary bivariate BM model
}
  \label{tab:table_5}
\end{table}

\begin{figure}[h!]
\centering\includegraphics[width=1\linewidth]{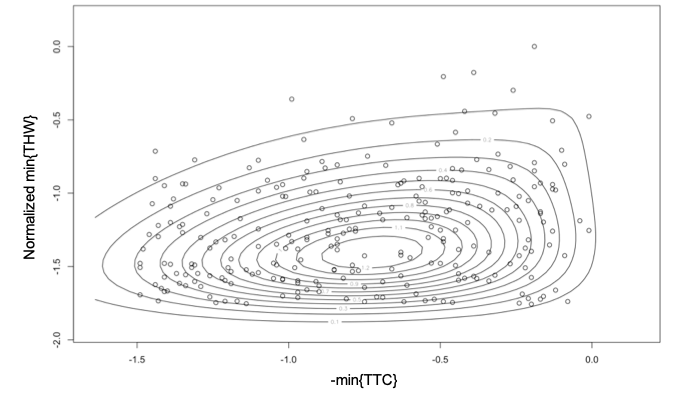}

\caption{Probability density contour plot for the stationary bivariate distribution and the observed data}
\label{img:figure_3}
\end{figure}

Other copula families were also tested to model the dependence between the negated values of TTC and of (normalized) THW using the range of families available in the \texttt{R} package \texttt{VineCopula}. To explore these families the marginals are estimated first (as per the previous section 4.1 of this paper), and the copula is here estimated subsequently. We first concluded that the copula with the better fit is the Joe-Frank copula \cite{brechmann2013cdvine}, with parameters 1.631 and 0.929. This result was confirmed by performing the goodness-of-fit test based on Kendall’s process (0.47 and 0.26 for the p-values of Cramer-von Mises statistic and Kolmogorov-Smirnov statistics, respectively). Simulating elements for the bivariate distribution analyzed in this exploratory approach, with a Joe-Frank copula and GEV (TTC) and Gumbel (THW) distributions for the margins, a maximum log likelihood of 50.73 and a Kendall’s tau of 0.184 are achieved. Using this fitted distribution, the estimated probability of having an accident, conditioned that both surrogate measures are below their threshold, is 0.0173 (0.0075, 0.0338), slightly higher and closer to the empirical compared to the bivariate logistic distribution estimation. For head-on collisions the obtained probability is 0.01300 (0.0048, 0.0280) and for rear-end collisions 0.0065 (0.00134, 0.01882).

The probability density function of this bivariate distribution with GEV margins is displayed in Figure 4. This contour plot provides the confidence of the regions for the empirical data points, showing the suitability of this model to estimate the joint probability of colliding with the opposite vehicle or with the passed vehicle. As previously mentioned, further analysis of this approach should be performed where approaches such as extreme copulas \cite{caperaa1997nonparametric} and the inclusion of other distributions for the margins should be explored.

The dependencies obtained for both the full ML estimation under the bivariate logistic distribution function \cite{gumbel1960distributions} or the step-based estimation of the Joe-Frank copula revealed the suitability of our approach to estimate the joint probability of an accident based on the two surrogate measures (TTC and THW). 

\begin{figure}[h!]
%\begin{subfigure}{}
  \centering
  \includegraphics[width=.5\linewidth]{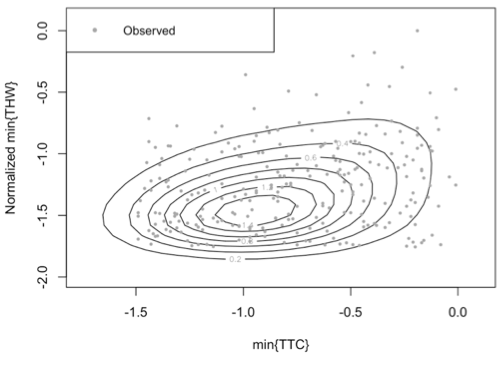}
  \label{fig:JFCopulaB}
%\end{subfigure}
%\begin{subfigure}{}
%  \centering
  \includegraphics[width=.5\linewidth]{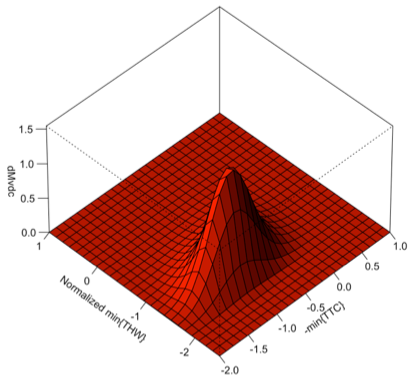}
  \label{fig:JFCopulaA}
%\end{subfigure}

\caption{a) Probability density contour plot for the stationary Joe-Frank Copula and the observed values (left), b) Probability density function for the stationary bivariate distribution}
\label{fig:JFCopula}
\end{figure}

As in the univariate case, we now aim at studying the importance of driver characteristics in the estimation of collision probability under the bivariate approach. Such analysis allows us to shed light on these variables in the interactions between the driver, the opposing and the passed vehicle. We consider again the covariates selected in the best univariate BM model for the head-on collisions and the selected stationary approach developed for the THW. A bivariate BM model was fitted to the joint distribution of max{-TTC} and (normalized) max{-THW}, achieving the parameters shown in Table \ref{tab:table_6}.
A tail dependence of 0.1302 and an estimated probability of accident of 0.0193 was obtained, much closer to the empirical probability.

\begin{table}[h]
\small
\centering
\begin{tabular}{l l l l}
\Xhline{2\arrayrulewidth}
 &  \textbf{TTC} &  \textbf{THW}\\
\Xhline{2\arrayrulewidth}
$\hat{\mu}_0$ & -1.949 (0.182) & -1.419 (0.147)	\\
$\hat{\mu}_1$\textit{(speedFront)} & 0.0326 \footnotesize{(0.0103)} & 0.00956 \footnotesize{(0.00810)}\\
$\hat{\mu}_2$\textit{(tailgatetp)} & -0.00489 \footnotesize{(0.00248)} & 0.00609 \footnotesize{(0.00221)}\\
$\hat{\mu}_3$\textit{(passinggap)} & -0.0240 \footnotesize{(0.00565)} & -0.0244 \footnotesize{(0.00544)}\\
$\hat{\mu}_4$\textit{(curvature)} & -33.4 \footnotesize{(18.96)} & 6.60 \footnotesize{(14.807)}\\
$\hat{\mu}_5$\textit{(speedpv)} & -0.00138 \footnotesize{(0.00506)} & 0.00606 \footnotesize{(0.0049)}\\
$\hat{\mu}_6$\textit{(Gender)} & -0.0462 \footnotesize{(0.0624)} & -0.141 \footnotesize{(0.04683)}\\
$\hat{\mu}_7$\textit{(Axious)}\&\textit{Hostile)} & -0.0342 \footnotesize{(0.0213)} & 0.0357 \footnotesize{(0.0164)}\\
$\hat{\sigma}$ & 0.395 \footnotesize{(0.023)} & 0.273 \footnotesize{(0.014)}\\
$\hat{\xi}$ &  -0.349 \footnotesize{(0.0657)} & -0.0651 \footnotesize{(0.0482)}\\
\Xhline{2\arrayrulewidth}
r	& 0.903 \footnotesize{(0.0365)}\\
AIC	& 392.3\\
\Xhline{2\arrayrulewidth}
\end{tabular}
\caption{Estimation results for the non-stationary bivariate BM model}
  \label{tab:table_6}
\end{table}

\section{Conclusions}
\label{S:4}

This paper analyzed the individual and joint probabilities of head-on collisions and rear-end collisions through the Block Maxima approach using the Univariate and Bivariate distributions to model dependence between the two surrogate measures that capture those types of collisions during passing maneuvers.
We investigated the fitting of EV models allowing for any real extreme value index, to understand how informative these models may be in respect to the extreme value indices pertaining to surrogate measures of safety. 
The univariate non-stationary estimation allowed to conclude that aspects linked to drivers’ characteristics, namely the gender, have a significant impact on the prediction of head-on collisions. However, these variables were not found to significantly improve the prediction of rear-end collisions regarding the stationary model. The bivariate model approach integrated the two different surrogate measures, TTC and THW, in order to estimate the risk of colliding with the opposite or with the passed vehicle in a single passing maneuver. Although the linear correlation between the two surrogate measures has proved to be weak, the bivariate distribution estimation shown the existence of dependency between these two surrogate measures and the exploration of further copula approaches seem to be promising.
Regarding the rear-end collisions model estimation, especially under small samples, we suggest also that fitting must concentrate on the Weibull distribution under BM approach, and GP distribution restricted to negative extreme value index under POT approach to improve precision on accident probability estimation in future work.
This exploratory analysis is the first attempt to explain how two different surrogate measures are linked, providing guidelines to estimate the probability of colliding with the opposite or passed vehicles even in the presence of weak correlation. 
To sustain the preliminary conclusions that both TTC and THW are good surrogate safety measures for near-accidents, head-on collisions and/or rear-end collisions, further analysis should be developed in order to validate through simulated data and/or data from other experimental scenarios the conclusions drawn by these models. Finally, the proposed probabilistic surrogate safety models should be integrated in traffic microscopic simulation frameworks for promising safety assessment, where the estimation of safety for individual maneuvers would not need to rely on accident records nor on the limitations of simulation‘s premise of accident-free models.

\bibliographystyle{model1-num-names}
\bibliography{sample.bib}
\bibliographystyle{apalike}


\appendix
\section{The POT Approach}
\label{appendix A}

According to the GP distribution an observation is identified as an extreme if it exceeds a predetermined threshold. The cumulative distribution function of exceedances $X$ over a threshold u (so called the conditional excess distribution function) is:

\begin{equation}
\label{eq:app_1}
F_u(y)=Pr\left(X-u \leq y | X>u \right), 0 \leq y \leq x_F -u
\end{equation}

Where $X$ is a random variable, u is a given threshold, $y=x-u$ are the excesses, and $x_F$ is the right endpoint of F. With a high enough threshold u, the conditional distribution $F_u (y)$ can be approximated by a GP distribution. The function of GP is given as follows:

\begin{equation}
\label{eq:app_1}
G(y)=1- \left[1+ \frac{\varepsilon\cdot y}{\sigma(z)} \right]^{\frac{-1}{\varepsilon}}
\end{equation}

where $\sigma(z)>0$ is the scale and $-\infty<\varepsilon<\infty$ is the shape parameter respectively.
The determination of the threshold in the POT approach determines the sample size. Therefore, an optimal threshold should be chosen so that the observations that exceed the threshold are real extremes, but still constitute a reasonable sample with relatively small variance. Choosing a small threshold will bias the results by considering normal observations as extremes, while choosing a high threshold would result with a few observations as extremes and thus large variability which would also bias the estimation results of the distribution.
In a non-stationary POT model, several factors (the vector z) can be included in the scale parameter to account for their impact on the probability of the extreme events. The POT method can also be used to study minima by considering the maxima of the negated values instead of minima of the original values. This is typically the case when applied for accident prediction using surrogate safety measures.

\section{Probability of Collision}
\label{appendix B}

The maximum domain of attraction condition holds if, for $\{X_1,...X_n\}$ i.i.d. sample of the surrogate measure of safety $x$,

%\begin{equation}
%\label{eq:app_2}
%\exists_{a_n >0}, b_n \in \mathbf{R}: \lim_{n \to \infty } P \left( \frac{\max_{i \leq n} x_i - b_n}{a_n} \leq y \right)= e^{- \left(1+ \frac{x_i (y-u)}{\sigma} \right)^{-\frac{1}{\varepsilon}}}, \varepsilon \in \mathbf{R}
%\end{equation}

\begin{equation}
\label{eq:app_2}
\exists_{a_n>0,\,b_n\in\mathbb{R}}:\lim_{n\to\infty}P\left(\frac{\max_{i\leq n}X_i-b_n}{a_n}\leq y\right)=e^{-\left( 1+\xi\frac{y-u}{\sigma}\right)_+^{-1/\xi}}
\end{equation}

\noindent with %$\left(1+ \frac{x_i (y-u)}{\sigma} \right)$
$1+\xi\frac{y-u}{\sigma}$ always positive. This formulation simplifies to a standard EV in the limit if $a_n$ and $b_n$ are chosen appropriately. Hence under the BM approach we define:

%\begin{equation}
%\label{eq:app_3}
%P_{BM} \left( collision \right) = \lim_{n \to \infty} P \left(\frac{\max_{i \leq n} x_i -b_n}{a_n} \leq 0 \right)
%= \lim_{n \to \infty} P \left(\max_{i \leq n} x_i -b_n \leq b_n  \right)
%\end{equation}

\begin{equation}
\label{eq:app_3}
P_{BM}(collision)=\lim_{n\to\infty}P\left(\frac{\max_{i\leq n}X_i-b_n}{a_n}\leq 0\right)=\lim_{n\to\infty}P\left(\max_{i\leq n}X_i\leq b_n\right)
\end{equation}

\noindent and estimate it by,

\begin{equation}
\label{eq:app_4}
\hat{P}_{BM} \left( collision \right) = e^{-\left(1- \hat{\xi} \frac{\hat{u}}{\hat{\sigma}}\right)^{-\frac{1}{\hat{\xi}}}}
\end{equation}

The above equation can also be interpreted as estimating the probability until achieving that “least real value” of the surrogate safety measure before collision, or approximately the probability of this measure being as close to zero up to the actual observed sample maximum.
In the POT approach the interpretation is similar but requires some adaptations. The EV condition rephrased in terms of POT condition leads to the GP distribution in the limit. That is, with $\tilde{x}$ the negated surrogate measure and $x^*$ the right end-point of its d.f.,

\begin{equation}
\label{eq:app_5}
\exists_{c_t>0}, d_t \in \mathbf{R}: \lim_{t \to x^*} P \left( \frac{\tilde{x} - d_t}{c_t} > y \arrowvert \tilde{x} > d_{t} \right) = (1+ \xi y )^{\frac{-1}{\xi}}
\end{equation}

As in BM the normalization constants can be chosen appropriately so for simplicity we write in its standard form.
Then define in this case, with $x_{n,n}$   representing the negated observed sample minima,

\begin{equation}
\label{eq:app_6}
P_{POT} \left(collision \right) = \lim_{t \to x^*} P \left( 
\frac{\tilde{x}-d_t}{c_t} > \frac{x_{n,n}-d_t}{c_t} \arrowvert \tilde{x} > d_t
\right)
\end{equation}

and estimate it by,

\begin{equation}
\label{eq:app_7 }
\hat{P}_{POT} \left(collision \right) = \left(1 + \frac{\hat{\xi}(x_{n,n}-\hat{d}_t)}{\hat{c}_t}
\right)^{-\frac{1}{\hat{\xi}}}
\end{equation}

\section{Summary of the Stationary Model Estimation}
\label{appendix C}

To validate the TTC and THW filter used in Section 4, a sensitivity analysis was performed. This analysis allows not only to analyze the stability of the estimation in the selection of the filter, but also to explore the range of shape parameter values to be expected in for non-stationary estimation. It also allows to evaluate the prediction performance of a simplified stationary approach. As presented in Appendix \ref{appendix B}, we also computed the estimates for the normalized model with $\{a_n, b_n\} = \{1-min(X_i)\}$ with $i \in 1, \cdot, n$ for the BM and $\{c_t, d_t\}= \{1-min(X_i)\}$ with $i \in 1, \cdot, t$ for the POT and study it’s performance compared to the non-normalized model. Such normalization can be of particular relevance under small samples and increased variability of the observed variables.

\begin{figure}[h!]
\centering\includegraphics[width=0.8\linewidth]{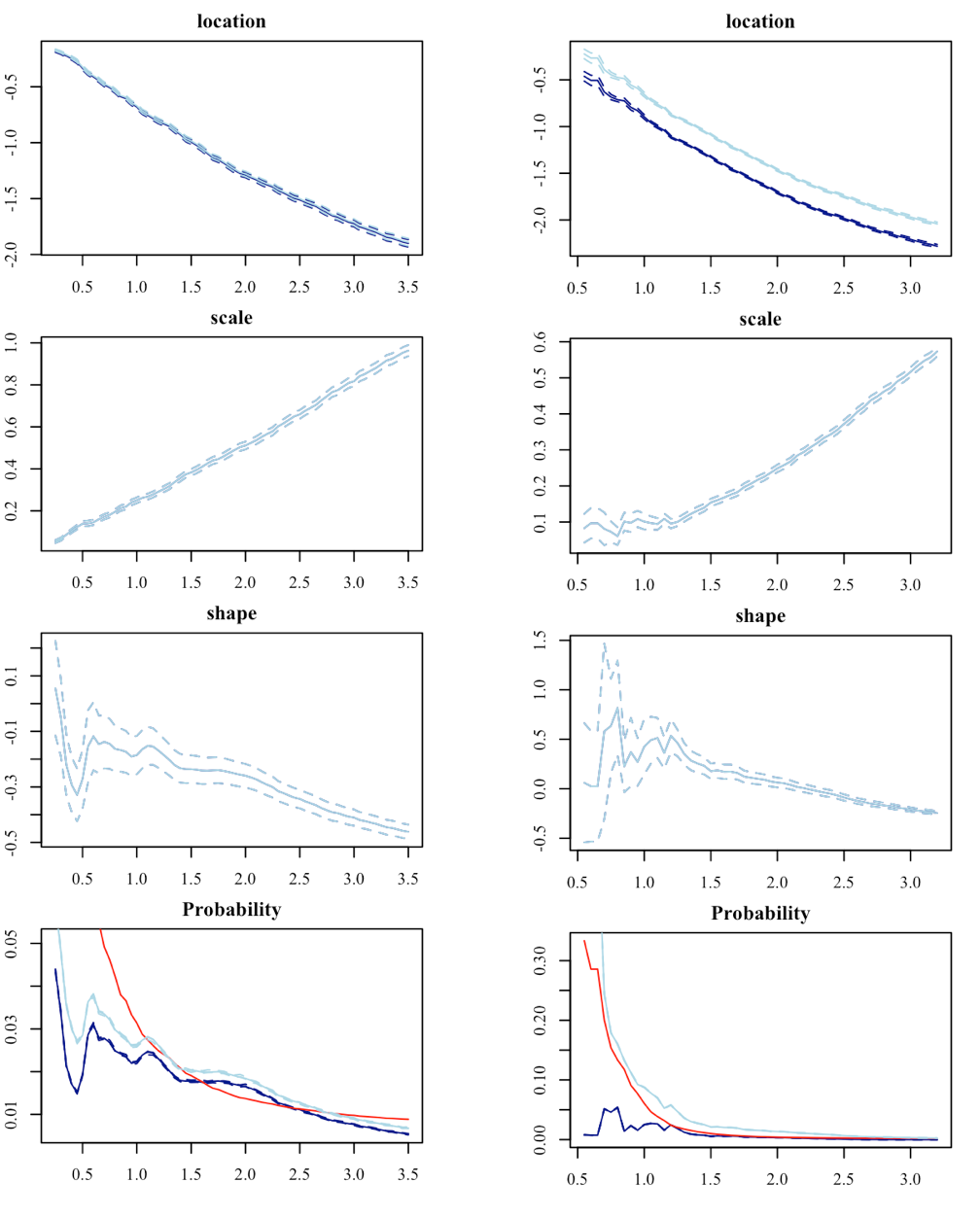}
\caption{ML estimates of the univariate stationary BM location, scale and shape parameters and the resulting probability of collision for the TTC (left) and the THW (right) for the original (dark blue) and normalized (light blue) model for different filtering criteria. In red is the empirical conditional collision probability.}
\label{img:figure_C1}
\end{figure}

In Figure \ref{img:figure_C1} we present the results for BM approach. The red curve in the bottom plot represents the empirical conditional collision probability from the used data. We first highlight the stable region for the estimate of the shape parameter for TTCs between 0.6 and 1.5s (head-on collision). The normalized model does not provide significant prediction improvement nor different estimates (as the dark blue and light blue curves for parameter estimates are overlapping), likely due to a large number of surrogate observations. From the analysis values between 1.2 and 1.5 s are likely to provide both good statistical properties and prediction performances. The value of 1.5s was thus selected for further analysis due to its smaller variance.

A first stable region for the estimate of the shape parameter for THWs happens between 1.0 and 1.3s (rear-end collision). However, within this region the shape parameter has positives values (between 0.35 and 0.55) therefore conflicting with the theoretical model. These results are mainly due to the small sample (<100 observations) and the high variability of THW within this region. Thus, we further explore the stable region between 2.0 and 2.4s where the shape parameter has values close to zero (between 0.06 and -0.05). The Gumbel distribution (GEV with $\varepsilon$=0) was tested within this region and used for further analysis. Finally, due to limitations of the THW dataset mentioned above, it is worth noting the increased prediction performance of the normalized model. 

In Figure \ref{img:figure_C2} we present the results for POT approach presented in Appendix \ref{appendix A}. This exploration helped in comparing and deciding between the BM and the POT approaches. Again, the red curve in the bottom plot represents the empirical conditional collision probability from the used data. We first highlight the stable region for the estimate of the shape parameter for TTC between 1.3 and 1.8s (head-on collision). The normalized model does provide prediction improvement, yet resulting probability estimates far from the empirical.

\begin{figure}[h!]
\centering\includegraphics[width=0.8\linewidth]{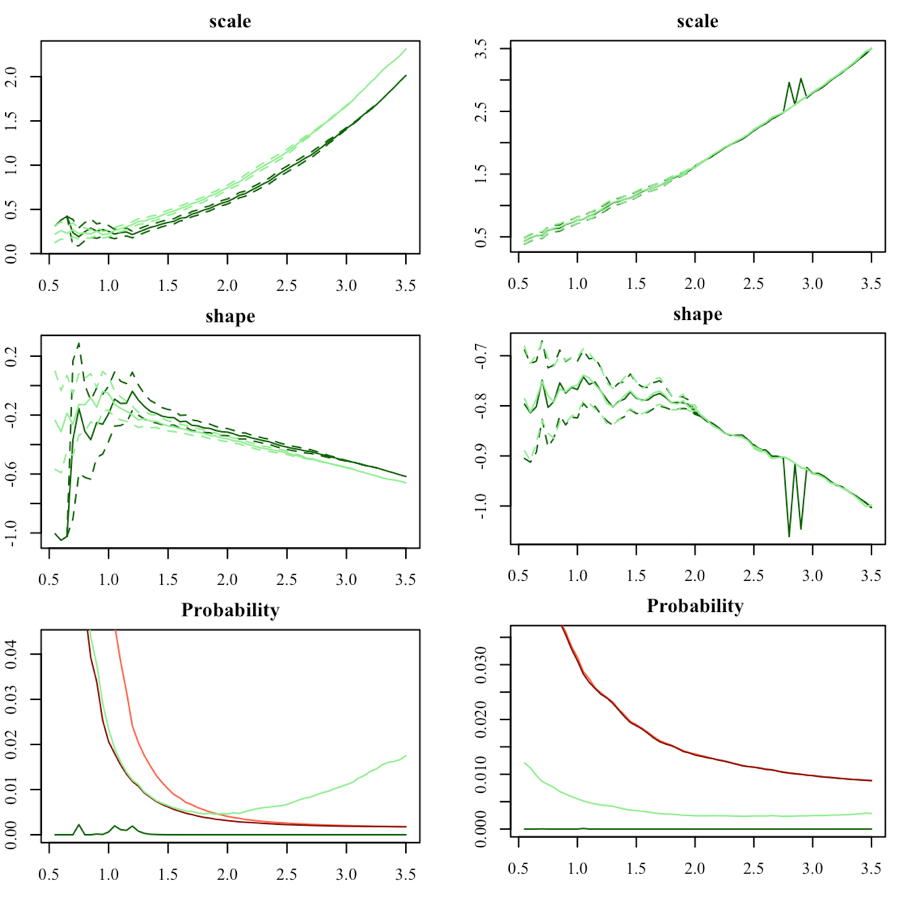}
\caption{ML estimates of the univariate stationary POT (a) scale and (b) shape parameters and the (c) resulting probability of collision for the TTC (left) and the THW (right) for the original (dark green) and normalized model (light green) for different threshold criteria. In red is the empirical conditional collision probability.}
\label{img:figure_C2}
\end{figure}

The stable region for the estimate of the shape parameter for THW happens between 1.2 and 2.0s (rear-end collision) with an estimate around -0.2. Similarly to the BM results, the normalized model does provide prediction improvement achieving predictions close to the empirical ones.


%% References


%\bibliographystyle{plainnat}





% \end{thebibliography}

\end{document}